\documentstyle[12pt]{article}

\begin{document}
\thispagestyle{empty}
\begin{center}
\LARGE \tt \bf{Cosmic Strings and Closed time-like curves in teleparallel gravity}
\end{center}
\vspace{2.5cm}
\begin{center} {\large L.C. Garcia de Andrade\footnote{Departamento de
F\'{\i}sica Te\'{o}rica - Instituto de F\'{\i}sica - UERJ
Rua S\~{a}o Fco. Xavier 524, Rio de Janeiro, RJ
Maracan\~{a}, CEP:20550-003 , Brasil.
E-mail : garcia@dft.if.uerj.br}}
\end{center}
\vspace{2.0cm}
\begin{abstract}
Closed Time-like curves (CTC) in Cosmic strings in teleparallel $T_{4}$ gravity are forbidden.This result shown here in $T_{4}$ was shown by Soleng (Phys.Rev.D49 (1994)1124) also to be valid in Einstein-Cartan (EC) gravity.Here we show that in $T_{4}$ to allow for CTC we are also led to a lower bound on the angular momentum of the cosmic string.This result is obtained by matching the interior $T_{4}$ solution to a General Relativity (GR) vacuum solution.One of the main differences of the present report and the one by Soleng is that here the interior symmetric solution does not have necessary polarized spins but only Cartan torsion in the spirit of teleparallelism.Torsion flux is computed and it is show that the center of cylinder singularity corresponds to a $2+1$ spacetime rotating point particle in $T_{4}$.Therefore the possibility of building time machines seems to be strongly constraint than in the case of EC gravity. 
\end{abstract}
\vspace{1.0cm}
\begin{center}
\end{center}
\newpage
\pagestyle{myheadings}
\markright{\underline{Cosmic strings and CTC in $T_{4}$}}
Recently H.Soleng and Letelier \cite{1,2,3,4} have investigated spinning cosmic string solutions in EC gravity.Letelier solution is easily shown to be trivial in Einstein teleparallelism \cite{5}.Soleng solution is based on the construction of a spin polarized cylinder which allows us to build a spinning cosmic string with torsion where only one component of torsion survives and where the spinning particles are direct along the infinite axis of the cylinder in Einstein-Cartan theory.Soleng showed based on this model that close to the cosmic spinning string (or near a point particle 
equations.In this report we show that in $T_{4}$ CTC curves would imply on a lower bound constraint on the cosmic string angular momentum in order to allow for CTC and consequently to build time-machines.Teleparallel solutions have recently attacted attention since the works of J.G.Pereira and his group \cite{6,7} who show that it is possible to explain old problems in physics such as the tetrad complex of gravitational energy by making use of teleparallelism \cite{6} as well as build some new teleparallel Scwarzschild and Kerr solutions \cite{7}and discussing the Lense-Thirring effect in teleparallel geometry.More recently yet Garcia de Andrade has shown \cite{8,9} that is possible to explain another problems in cosmology such as the global rotation of the universe by using G\"{o}del teleparallel solution in a very natural way.Besides as shown very recently it is possible to obtain a system of a cosmic string inside a black hole as a $T_{4}$ solution.Let us begin by considering the spacetime metric with cylindrical symmetry considered by Soleng as given by the one-form basis     
\begin{equation}
{\omega}^{0}=dt+Md{\phi}
\label{1}
\end{equation}
\begin{equation}
{\omega}^{1}=dr
\label{2}
\end{equation}
\begin{equation}
{\omega}^{2}={\rho}d{\phi}
\label{3}
\end{equation}
\begin{equation}
{\omega}^{3}=dz
\label{4}
\end{equation}
In terms of Cartan exterior differential forms the Soleng metric can be expressed as 
\begin{equation}
ds^{2}=({\omega}^{0})^{2}-({\omega}^{1})^{2}-({\omega}^{2})^{2}-({\omega}^{3})^{2}
\label{5}
\end{equation}
Now from the Cartan structure equations
\begin{equation}
Q^{i}=d{\omega}^{i}+{\omega}^{i}_{j}{\wedge}{\omega}^{j} 
\label{6}
\end{equation}
where ${\omega}^{i}_{j}$ is the connection one-form and
\begin{equation} 
R^{i}_{j}=R^{i}_{jkm}({\Gamma}){\omega}^{k}{\wedge}{\omega}^{m}=d{\omega}^{i}_{j}+{\omega}^{i}_{k}{\wedge}{\omega}^{k}_{j}
\label{7}
\end{equation}
Here ${\wedge}$ is the exterior product of forms symbol and $R^{i}_{jkl}({\Gamma})$ are the components of the Riemann-Cartan geometry curvature tensor and $R^{i}_{j}$ is the curvature $2-form$.Rewriting the metric (\ref{1}) in the differential forms language one obtains
\begin{equation}
ds^{2}={\eta}_{ij}{\omega}^{i}{\omega}^{j}
\label{8}
\end{equation}
where ${\eta}_{ij}=diag(+1,-1,-1,-1)$ is the tetrad Minkowski metric.By making use of the teleparallel condition $R^{i}_{jkl}({\Gamma})=0$ into the equation (\ref{7}) we notice that the constraint 
\begin{equation}
{\omega}^{i}_{j}=0
\label{9}
\end{equation}
fulfills the teleparallel condition.Here we addopt this stronger teleparallel condition which also has been addopted by Letelier \cite{5} in the construction of torsion loops in teleparallel spacetimes.By using the condition (\ref{9}) into equation (\ref{6}) one obtains the torsion 2-form in the form
\begin{equation}
Q^{i}=d{\omega}^{i}=T^{i}_{jk}{\omega}^{j}{\wedge}{\omega}^{k}
\label{10}
\end{equation}
where $T^{i}_{jk}$ are the components of the Cartan's torsion tensor.Applying this simple expression to the expressions for the basis one-forms of the cosmic spinning torsion string system above one obtains after a quick computation one obtains the components of the torsion tensor as
\begin{equation}
s_{0}=T^{0}_{12}=\frac{M_{,r}}{\rho} 
\label{11}
\end{equation}
\begin{equation}
s_{1}=T^{2}_{12}=\frac{{\rho}_{,r}}{\rho}
\label{12}
\end{equation}
By assuming that both torsion components inside the cylinder are constants
equations (\ref{11}) and (\ref{12}) become differential equations that can be solved immediatly to yield the interior solution 
\begin{equation}
ds^{2}=[dt+\frac{s_{0}}{s_{1}}e^{s_{1}r}d{\phi}]^{2}-[dr^{2}+dz^{2}]-e^{2s_{1}r}d{\phi}^{2}
\label{13}
\end{equation}
This metric has some very nice properties that now we wish to comment.First according to Paul Tod \cite{10} classification of the conical metric
\begin{equation}
ds^{2}=(dt+{\alpha}d{\phi})^{2}-{\beta}^{2}r^{2}d{\phi}^{2}-(dz+{\gamma}d{\phi})^{2}
\label{14}
\end{equation}
when the last term can be ommited but ${\alpha}$ is non-zero the metric would represented a point particle rotating in $2+1$ gravity.Therefore with this classification one could say by comparison with our interior solution (\ref{13}) one notices that this solution could be interpreted as a singular behaviour at $r=0$ where a point particle would rotate at the center of the cosmic string at $3+1$ gravity spacetime.Before making use of the matching conditions to disccuss CTC conditions we could compute the torsion flux by making use of geometrical phases formula
\begin{equation}
{\int}_{\Sigma}{Q^{0}}=\int{{\omega}^{0}}=2{\pi}\frac{s_{0}}{s_{1}}e^{s_{1}r}
\label{15}
\end{equation}
which shows that at the conic singularity $r=0$ torsion flux is constant.CTC could be immagined as torsion loops around the cosmic strings \cite{6}.Let us now pointed out that the interior solution to be physicsl should be matched to an exterior solution preferable with a GR vacuum of the type given by Soleng one-forms
\begin{equation}
{\theta}^{0}=dt+\frac{J}{2{\pi}}d{\phi}
\label{16}
\end{equation}
\begin{equation}
{\theta}^{1}=dr
\label{17}
\end{equation}
\begin{equation}
{\theta}^{2}=(1-\frac{{\mu}}{2{\pi}})(r+r_{0})d{\phi}
\label{18}
\end{equation}
\begin{equation}
{\theta}^{3}=dz
\label{19}
\end{equation}
where J is the intrinsic angular momentum of the cosmic string and ${\mu}$ is the string mass.The Arkuszewski-Kopczynski-Ponomariev (AKP) \cite{11} mactching conditions of EC theory can be used here \cite{2} 
\begin{equation}
g_{22}|_{+}=g_{22}|_{-}
\label{20}
\end{equation}
\begin{equation}
g_{02,1}|_{+}=g_{02,1}|_{-}-T_{012}
\label{21}
\end{equation}
\begin{equation}
g_{22,1}|_{+}=g_{22,1}|_{-}-2T_{212}
\label{22}
\end{equation}
where the plus sign refers to the exterior solution and the minus sign refers to the interior solution.Substitution of the respective above solutions into the matching conditions yields
\begin{equation}
e^{2s_{1}r}=-(1-\frac{\mu}{2{\pi}})^{2}(R+r_{0})^{2}
\label{23}
\end{equation}
\begin{equation}
2\frac{s_{0}}{s_{1}}e^{s_{1}R}=\frac{J}{\pi}-s_{0}
\label{24}
\end{equation}
where R is the radius of the cylinder.From these last two expressions allows us to determine the constant torsion components in terms of the cylinder radius.They also allows us to obtain a constraint between the angular momentum of the $T_{4}$ string and the ratio of the torsion components
\begin{equation}
{\pi}\frac{s_{0}}{s_{1}}=\frac{J}{[s_{1}+2e^{s_{1}r}]}
\label{25}
\end{equation}
Let us now from these expressions to examine the problem of construction of CTC in $T_{4}$.We know from GR that the CTC are allowed with the following constraint
\begin{equation}
(1-\frac{\mu}{2{\pi}})(r+r_{0})<\frac{J}{2{\pi}}
\label{26}
\end{equation}
which at the surface of the cylinder is
\begin{equation}
(1-\frac{\mu}{2{\pi}})(R+r_{0})<\frac{J}{2{\pi}}
\label{27}
\end{equation}
To simpland from the matching conditions we reduce this expression to
\begin{equation}
\frac{s_{0}}{[1+\sqrt{2\frac{s_{0}}{s_{1}}}]}<\frac{J}{2{\pi}}
\label{28}
\end{equation}This expression can be further simplified if we consider that torsion can be "isotropic" or that they have the same $s_{0}=S_{1}$To simpland from the matching conditions we reduce this expression to
\begin{equation}
\frac{s_{0}}{[1+\sqrt{2}]}<\frac{J}{2{\pi}}
\label{29}
\end{equation}
Nevertheless by asimilar reasoning used by Soleng we can show that the assuption that CTC can be constructed leads us to a contradiction.Let us assume that $R>>r_{0}$ and $\frac{\mu}{2{\pi}}<<<1$ thus from expression (\ref{26}) one obtains that  
\begin{equation}
R<\frac{J}{2{\pi}}
\label{30}
\end{equation}
Therefore if we consider that$R=\frac{h}{m}$ and $J={\mu}\frac{h}{m}$ one may conclude that ${\mu}>1$ which contradicts our assumption that ${\mu}{2{\pi}}<<1$.Therefore one may conclude that there is no possibility of building CTC in $T_{4}$ as in the EC case.
\section*{Acknowledgement}
I am very much indebt to Prof.P.S.Letelier, Prof.Harald Soleng,Prof.J.G.Pereira and Prof.I.D.Soares for helpful discussions on the subject of this paper.Financial support from CNPq. and UERJ is gratefully acknowledged.

\end{document}